\documentclass[doublecol]{epl2} 

\renewcommand{\d}{{\rm d}}
\newcommand{\mub}{\bar\mu}
\newcommand{\bi}{{\rm bi}}
\newcommand{\rs}{r_h}

\newcommand{\half}{{\frac{1}{2}}}

\newcommand{\myskip}[1]{}    
       
\newcommand{\p}{\partial}   
   
\newcommand{\twodots}{\,\,\cdot\,\,\cdot\,}   
   
\newcommand{\ed}{\cdot}
\newcommand{\td}{\twodots}   
      
\newcommand{\Om}{\Omega}
\newcommand{\BEQ}{\begin{eqnarray}}   
\newcommand{\EEQ}{\end{eqnarray}}   
\newcommand{\BEA}{\begin{eqnarray}}   
\newcommand{\EEA}{\end{eqnarray}}   
\newcommand{\nn}{\nonumber }   
\renewcommand{\d}{{\rm d}}

\newcommand{\mn}{{\mu\nu}}

\title{Einstein vs Maxwell: Is gravitation a curvature of space, \\
a field in flat space, or both?}
\shorttitle{Einstein vs Maxwell}
\author{Theo M. Nieuwenhuizen}

\shortauthor{Th.M. Nieuwenhuizen}

\institute{Institute for Theoretical Physics,  University of Amsterdam,   
Valckenierstraat 65, 1018 XE Amsterdam, The Netherlands}   

\abstract{
Starting with a field theoretic approach in Minkowski space, 
the gravitational energy momentum tensor is derived  from the Einstein equations 
in a straightforward manner. This allows to present them as  
{\it acceleration tensor} = const. $\times$ {\it total energy momentum tensor}.
For flat space cosmology the gravitational energy is negative and cancels the material energy.
In the relativistic theory of gravitation a bimetric coupling between the Riemann and
Minkowski metrics breaks general coordinate invariance.
The case of a positive cosmological constant is considered.
A singularity free version of the Schwarzschild black hole is solved analytically.
In the interior the components of the metric tensor quickly die out, 
but do not change sign, leaving the role of time as usual.
For cosmology the $\Lambda$CDM model is covered, while there appears a form of inflation 
at early times. Here both the total energy and the zero point energy vanish.}

\begin{document}

\maketitle

It is said that in introducing the general theory of relativity (GTR), 
Einstein made the step that Lorentz and Poincar\'e had failed to make: 
to go from flat space to curved space. Technically, this arises from the
group of general coordinate transformations~\cite{elephant,WeinbergGR}.
One fundamental difficulty is then how to deal with the physics of gravitation itself,
since there is only a quasi energy-momentum tensor~\cite{LandauLifshitz}. 
For gravitational wave detection, e.g., this leaves open the question as to 
how energy can be faithfully transferred from the wave to the detector.
The proper energy momentum tensor of gravitation was derived only recently 
by Babak and Grishchuk~\cite{BabakGrishchuk}, who start 
with a field theoretic approach to gravitation, 
in terms of a tensor field $ h^\mn$ in a Minkowski background space-time. 
The metric of the latter, $\eta_\mn={\rm diag}(1,-1,-1,-1)$, 
is denoted in arbitrary coordinates by $\gamma_\mn=(\gamma^\mn)^{-1}$. 
The Riemann metric tensor $g_\mn=(g^\mn)^{-1}$, is then defined by
\BEQ\label{gMNphiMN=} 
\sqrt{\frac{g}{\gamma}}g^\mn=\gamma^\mn+ h^\mn\equiv k^\mn,
\qquad \frac{g}{\gamma}=\frac{{\rm det}(g_\mn)}{{\rm det}(\gamma_\mn)}.
\EEQ 
It is just a way to code the gravitational field,
allowing to expresses distances by $\d s^2=g_\mn\d x^\mu\d x^\nu$.
Such a non-linear way to code distances in a flat space is not uncommon.
For diffuse light transport through clouds, one may express distances 
in the optical thickness, the number of extinction lengths. 
If the cloud is not homogeneous, points at the same physical distance 
are described by a different optical distance and, vice versa.

The Maxwell view that gravitation is a field in flat space, 
was actually the starting point for Einstein, and reappeared regularly.
Nathan Rosen~\cite{Rosen}, coauthor of the Einstein-Podolsky-Rosen paper
that led the basis for quantum information, considers a bimetric theory, 
involving  the Minkowski metric and the Riemann metric. 
Bimetrism is quite natural, with $\eta_\mn$ entering e.g. particle physics, 
and $g_\mn$ e.g. cosmology.
Rosen considers covariant derivatives $D_\mu$ of Minkowski space,
with Christoffel symbols $\gamma^\lambda_{\ed\mn}$ vanishing in Cartesian coordinates.
When replacing 
in the Riemann Christoffel symbols partial derivatives by Minkowski covariant ones,
\BEA 
\Gamma^\lambda_{\ed\mn}&=&\half g^{\lambda\sigma}(\p_\mu g_{\nu\sigma}+\p_\nu g_{\mu\sigma}-
\p_\sigma g_{\mn})
\mapsto \nn\\
 G^\lambda_{\ed\mn}&=&\half g^{\lambda\sigma}(D_\mu g_{\nu\sigma}+D_\nu g_{\mu\sigma}-
D_\sigma g_{\mn}), \EEA
the obtained Christoffel-type symbols $ G^\lambda_{\ed\mn}$ are tensors in Minkowski space. 
Inspired by the Landau-Lifshitz and Babak-Grishchuk results, we may define the 
{\it acceleration tensor}
\BEQ \label{AMN=}
A^\mn=\half D_\alpha D_\beta(k^\mn k^{\alpha\beta}-k^{\mu\alpha}k^{\nu\beta}),\quad
\EEQ 
where $k^\mn=\gamma^\mn+h^\mn$ and 
in which the $\gamma\gamma$ terms do not contribute. Then we can calculate the combination
\BEQ \label{tauMN}
\tau^\mn=\frac{c^4}{8\pi G}\left[\frac{\gamma}{g}A^\mn-(R^\mn-\half g^\mn R)\right].\EEQ 
In doing so, we make use of Rosen's observation that $R^\mn$ remain unchanged if one
replaces all partial derivatives by covariant ones in Minkowski space~\cite{Rosen}.
It appears that all second order derivatives drop out from (\ref{tauMN}), leaving
a bilinear form in first order covariant derivatives, 

\BEQ \label{tauMN=}
&& \tau^\mn=
\frac{c^4\,\gamma}{8\pi G\,g}\left(
\half h^\mn_{\td:\lambda}h^{\lambda\rho}_{\td:\rho}
-\half h^{\mu\lambda}_{\td:\lambda}h^{\nu\rho}_{\td:\rho}\right. 
\nn\\&&+\half h^{\mu\lambda:\rho}h^\nu_{\ed\lambda:\rho} 
+\frac{1}{4}k^\mn h^{\lambda\rho:\sigma}h_{\lambda\sigma:\rho}
-\half h^{\lambda\rho:\mu}h^\nu_{\ed\lambda:\rho}
\nn\\&&-
\half h^{\mu\lambda:\rho}h^{\td:\nu}_{\lambda\rho}
+\frac{1}{4}h^{\lambda\rho:\mu}h^{\td:\nu}_{\lambda\rho}
-\frac{1}{8}h_\lambda^{\ed\lambda:\mu}h_\rho^{\ed\rho:\nu}\nn\\&&
-\frac{1}{8}k^\mn h^{\lambda\rho:\sigma}h_{\lambda\rho:\sigma}
+\left.\frac{1}{16}k^\mn h_\rho^{\,\ed\rho:\lambda}h^\sigma_{\,\ed\sigma:\lambda} 
\right). \EEQ
in which
$X_{:\mu}\equiv D_\mu X$ and raising (lowering) of indices of $h^\mn_{\td:\rho}$ 
is performed with $k^\mn$ ($k_\mn$). $\tau^\mn$ is a tensor in Minkowski space.
For Cartesian coordinates, it coincides with the Landau-Lifshitz quasi-tensor.
In general, it coincides with the Babak-Grishchuk tensor $\gamma t^\mn/g$.
Inclusion of matter is now much easier than in ~\cite{BabakGrishchuk}. Inserting
the Einstein equations in the right hand side of (\ref{tauMN}), we may write the 
Einstein equations in the Newton shape: acceleration=mass$^{-1}\times$force,
\BEA \label{MaxEinEq=}
A^\mn&=&\frac{8\pi G}{c^4}\Theta^\mn,\nn \\ 
\Theta^\mn&=&\frac{g}{\gamma}\theta^\mn,\qquad
\quad \theta^\mn\equiv \tau^\mn+T^\mn.\EEA 
$\Theta^\mn$ is {\it the total energy momentum tensor of gravitation and matter}.
It is conserved, $D_\nu \Theta^\mn=0$, since  Eq. (\ref{AMN=}) implies 
$D_\nu A^\mn=0$, because covariant Minkowski derivatives commute.

As an application, let us consider cosmology, described by
the Friedman-Lemaitre-Robertson-Walker (FLRW) metric, 

\BEQ \label{metric=}
\d s^2&=&U(t)c^2\d t^2-V(t)\left(\frac{\d r^2}{1-kr^2}+r^2\d\Omega^2\right), \\
\d\Omega^2&=&\d\theta^2+\sin^2\theta\d \phi^2.\nn
\EEQ
Let us consider flat space, $k=0$, and $U=1$, $V(t)=a^2(t)$ with $a$ the scale factor.
Then $\d s^2=c^2\d t^2-a^2(t)\d{\bf r}^2$ is space-independent, implying that 
$A^{00}=0$, due to the shape (\ref{AMN=}). According to (\ref{MaxEinEq=}) it then
follows that {\it the total energy density is zero}, because {\it the gravitational
energy density}, $\tau^{00}=-3c^4\dot a^2/(8\pi Ga^2)$, {\it is negative and cancels the
one of matter}, $T^{00}=\rho$,  due to the Friedman equation.
In other words, {\it such a universe contains no overall energy}.

So far we have discussed an alternative, field theoretic formulation of GTR. 
If we consider a local energy momentum density as a {\it sine qua non} property,
then we are led to consider Minkowski space as a fixed ``pre-space'', that
exist already without matter, just as a region of space ahead of the earth's orbit is
right now almost empty (Minkowskian), and when the earth arrives, there will
be more gravitational and matter fields, but, in our view, no change of space.
Also for cosmology there is a different interpretation. In GTR coordinates are
fixed to clusters of galaxies, this is called ``coordinate space'', but due to the 
increasing scale factor galaxies are said to move away from each other:
physical space (i.e. Riemann space) is said to expand. Here we are led to another view:
Coordinate space is physical space, so clusters of galaxies do not move 
away from each other in time.
\cite{LogunovBook}
However, the cosmic speed of light $\d r/\d t=c/a(t)$,
which was very large at early times, 
keeps on decreasing, thus causing a redshift, till $a$ is infinite,
when galaxies are invisible.

{\it  Relativistic Theory of Gravitation, RTG}.
Let us move on to an extension of GTR, giving up general coordinate invariance.
Discarding a total derivative of the Hilbert-Einstein action, 
Rosen expresses the gravitational action $S_R=\int\d^3x\d t\sqrt{-g\,}\,L_R$
in terms of 
~\cite{Rosen}
\BEQ \label{LR1=}
L_R&=&
\frac{c^4g^\mn}{16\pi G}(G^\lambda_{\ed\mn} G^\sigma_{\ed\lambda\sigma}
-G^\lambda_{\ed\mu\sigma}G^\sigma_{\ed\nu\lambda})=\frac{c^4\sqrt{\gamma/g}}{128\pi G} \\
&\times&
(2h^{\mu\nu:\rho}h_{\mu\nu:\rho}-4h^{\mu\nu:\rho}h_{\mu\rho:\nu}
-h^{\nu}_{\,\ed\nu:\mu}h_\rho^{\,\ed\rho:\mu}).\nn
\EEQ
Involving only Minkowski covariant first order derivatives, 
it is close to general approaches in field theory. 
Logunov and coworkers continue on this~\cite{LogunovBook}. 
The subgroup of gauge transformations that transform $ h^\mn$ but leave
coordinates invariant, allows three extra terms
~\cite{LogunovBook},
\BEQ \label{Lg1=}
L_g&=&L_R
-\rho_\Lambda+\half\rho_\bi \gamma_\mn g^\mn-\rho_0\sqrt{\gamma/g}.
 \EEQ
Here $\rho_\Lambda$ is the familiar energy related to a cosmological constant.
The $\rho_0$ term describes a harmless shift of the zero level of energy, 
$\delta S=-\int\d^3x\d t\sqrt{-\gamma}\rho_0$. 
The bimetric term $\rho_\bi $ couples the Minkowski and the Riemann metrics.
It acts like a mass term, because it breaks general coordinate invariance,
and has some analogy to a mass term in massive electrodynamics. 
Logunov then imposes the relation
\BEQ \label{omomom}\rho_\Lambda=\rho_\bi=\rho_0 ,\EEQ
which, in the absence of matter, keeps space flat,
$ h^\mn=0$, $g^\mn=\gamma^\mn$ and also $L_g=0$.
Thus one free parameter remains.
Logunov's choice $\rho_\bi \equiv -m^2c^4/(16\pi G)<0$ leads to an inverse length $m$ and, 
in quantum language, a graviton mass $\hbar m/c$.
The negative cosmological constant can be counteracted by an inflaton field ~\cite{LogInflaton}.
The obtained theory has some drawbacks, such as self-repulsive properties for matter 
falling onto a black hole, and a minimal and a maximal size of the scale factor in cosmology
~\cite{LogunovBook}\cite{LogInflaton}. For a related approach to finite range 
gravity, based on a generalized Fierz-Pauli coupling, see \cite{BabakGrishchuk2}.

We shall focus on the opposite choice, a positive cosmological constant $\Lambda$, 
~\cite{ThemAgain}\cite{WMAP}
\BEQ \label{rhobi=rhoLambda}
\rho_\Lambda&\equiv&
\frac{\Lambda c^4}{8\pi G}=\frac{3c^2}{8\pi G}\,\Om_{v,0}H_0^2
=\frac{3c^2}{8\pi G}\,0.74\left(\frac{0.71}{9.78 \textrm{Gyr}}\right)^2, \nn\\
\rho_\bi&\equiv&\frac{\Lambda_\bi c^4}{8\pi G}=\rho_\Lambda.
\EEQ 
Now the graviton has an ``imaginary mass'', $m=\hbar\sqrt{-2\Lambda_\bi}/c$, 
it is a ``tachyon'': Gravitational waves are unstable at today's Hubble scale.
But this is of no concern, since on that scale, not single gravitational waves
but the whole Universe matters, being unstable (expanding) anyhow.

Though we take $\rho_\bi= \rho_\Lambda$, $\Lambda_\bi=\Lambda$, 
our further notation is valid for the general case
$\rho_\bi\neq \rho_\Lambda$,  $\Lambda_\bi\neq\Lambda$.

The Einstein equations that couple the Riemann metric to matter 
read
\BEQ \label{Einst=} 
&&R^\mn-\half g^\mn R=\frac{8\pi G}{c^4} T^\mn_{\rm tot},\qquad \\
&&T^\mn_{\rm tot}=T^\mn+\rho_\Lambda g^\mn+\rho_\bi\gamma_{\rho\sigma} (
g^{\mu\rho}g^{\sigma\nu}-\half g^\mn g^{\rho\sigma}).\nn
\EEQ
Conservation of energy momentum, $T^\mn_{{\rm tot};\nu}=0$,
imposes a constraint due to the $\rho_\bi$ terms,~\cite{LogunovBook}
\BEQ \label{flatspacecond}
D_\nu\left(\sqrt{\frac{g}{\gamma}}g^\mn\right)=0,\qquad \textrm{or}\qquad D_\nu h^\mn=0,
\EEQ  
which for Cartesian coordinates coincides with the GTR harmonic condition 
$\p_\nu( \sqrt{-g}g^\mn)=0$ ~\cite{WeinbergGR}.
Thus the theory automatically demands the harmonic constraint for $g^\mn$, or, 
equivalently, the Lorentz gauge for $h^\mn$, thereby severely reducing 
the gauge invariance of GTR.
  
Changes of Einstein's GTR have mostly met deep troubles with one or another established
property, though not all proposals are ruled out~\cite{elephant,CliffordWill}.
The present one is rather subtle and promising. For most applications, 
the Hubble-size $\rho_\Lambda=\rho_\bi$ terms in Eq. (\ref{rhobi=rhoLambda},\ref{Einst=}) 
are too small to be relevant, so known results from general relativity can be reproduced. 
Indeed, viewed from a GTR standpoint, Eq. (\ref{flatspacecond}) is only a particular gauge, 
and actually often considered, while the cosmological constant only plays a role in cosmology.
Logunov checked a number of effects in the 
solar system: deflection of light rays by the sun, the delay of a radio signal, the 
shift of Mercury's perihelion, the precession of a gyroscope, and the gravitational 
shift of spectral lines.~\cite{LogunovBook}
Likewise, we expect agreement for binary pulsars.~\cite{CliffordWill}
Differences between GTR and RTG may arise, though, for large gravitational fields,
that we consider now.

{\it Black holes}. It is known that true black holes, objects that have a 
horizon, do not occur in the RTG with $\rho_\Lambda,\rho_\bi\to0$.~\cite{Rosen} 
But there are solutions very similar to it, that might be named ``grey holes'', 
but we just call them ``black holes''.
The Minkowski line element in spherical coordinates is simply
$\gamma_\mn\d x^\mu\d x^\nu=c^2\d t^2-\d r^2-r^2\d\Omega^2$.
The one of Riemann space is 
\BEQ \d s^2=g_\mn\d x^\mu\d x^\nu=U(r)c^2\d t^2-V(r)\d r^2-W^2(r)\d\Omega^2.\EEQ 
In harmonic coordinates, the Schwarzschild black hole is described by~\cite{WeinbergGR}
\BEQ \label{UVWSs} U_s=\frac{1}{V_s}=\frac{r-\rs}{r+\rs},\quad 
W_s=r+\rs,\quad \rs=\frac{GM}{c^2}. \EEQ
The horizon radius $r_h$ equals half the Schwarzschild radius.
Let us scale $r\to r\rs$, and define
\BEQ \label{Uexpu} U=e^u,\quad V=e^v,\quad W=2\rs e^w, \EEQ
so that $w$ is small near the horizon. 
The dimensionless small parameter arising from $\rho_\bi=\rho_\Lambda$, is very small,
\BEQ \label{mub=} \bar\lambda\equiv r_h\sqrt{2\Lambda}=
2.38\, 10^{-23}\,\frac{M}{M_\odot}.
\qquad \bar\mu\equiv r_h\sqrt{2\Lambda_\bi}=\bar\lambda.
\EEQ   

The sum and difference of the $(t,t)$ and $(r,r)$ 
Einstein equations give
\BEQ\label{Einp} 
&{}&\frac{1}{2}e^{v-2w}-w'(u'-v'+4w')-2w''\nn \\
&{}&=e^v(\bar\lambda^2-\frac{1}{4}\mub^2r^2e^{-2w})+\frac{8\pi Gr_h^2}{c^4} e^v(\rho-p), \\
\label{Einm} 
&{}& w'(u'+v'-2w')-2w''\nn\\
&&=\frac{1}{2}\mub^2(e^{v-u}-1)+\frac{8\pi Gr_h^2}{c^4} e^v(\rho+p),
\EEQ 
respectively. The harmonic condition imposes 
\BEQ \label{Eino} u'-v'+4w'=r\exp({v-2w}). \nn\EEQ

In the Schwarzschild black hole of GTR, there is no matter outside the origin.
We shall focus on that situation. A parametric solution of these equations then reads
\BEQ \label{solr}
r&=&\frac{1+\eta(e^\xi+\xi+\log\eta+r_0)}{1-\eta(e^\xi+\xi+\log\eta+r_0)}, 
\qquad \label{solu}\\
 u&=&\xi+\log\eta,\nn
\\
\label{solv} v&=&\xi-\ln\eta-2\log(e^\xi+1),
\qquad \\
\label{solw} w&=&\eta e^\xi+\mub^2(\xi+\log\eta+w_0). \nn
\EEQ
where $\xi$ is the running variable and $\eta$ is a small scale.
Corrections of next order in $\eta$ can be expressed in 
dilogarithms, but they are not needed since $\mub$ is very small.

To fix the scale $\eta$, we note that energy momentum conservation implies,
 as in GTR, $(\rho+p)u'+2p'=0$.
In the stationary state all matter is located at the origin, 
which is only possible if $p(r)\equiv 0$, implying $\rho(r)u'(r)=0$.
This is obeyed for $r\neq0$ since $\rho=0$ there, but since $\rho(0)>0$ 
(it is infinite), we have to demand $u'(0)=0$. 
Let us define a factor $\alpha$ by
$\alpha={\mub^2}/{\eta}$. The above solution brings
$w'(r)={\p_\xi w}/{\p_\xi r}=({e^\xi+\alpha})/[{2(e^\xi+1)}],$
so in the interior $w'=\half\alpha$.
Since $e^{v}\ll 1$ there, Eqs. (\ref{Einp},\ref{Einm}) 
confirm that $ w''=0$, and with
$w(1)={\cal O}(\eta)$ this solves $w(r)=\half\alpha\,(r-1)$.
Moreover, from  the harmonic constraint 
(\ref{Eino}) we have  in the interior
$u(r)-v(r)+4w(r)={\rm const}=2\ln\eta$,
implying that Eq. (\ref{Einm}) yields in the interior
$u'(r)=\{\exp[2\alpha(r-1)]-\eta^2-\mub^2\}/(2\eta)$.
From $ u'(0)=0$ we can now solve $\alpha$,
\BEQ \label{alpha=}\alpha=\log\frac{1}{\mub},\qquad \eta=\frac{\mub^2}{\ln1/\mub}. 
\EEQ

As seen in fig. 1, our solution (\ref{Uexpu},\ref{solu},-\ref{alpha=}) 
coincides with Schwarzschild's for $\xi\gg1$.
In the regime $\xi={\cal O}(1)$, there is a transition
towards the interior $\xi\ll-1$, where exponential corrections 
can be neglected. Both $U=\eta e^\xi$ and $V=e^\xi/\eta$ are very small
there, but, contrary to the Schwarzschild case, they remain positive:
{\it The behavior in the interior of the RTG black hole is not qualitatively 
different from usual, be it that the gravitational field is large}.

\begin{figure}
\onefigure[width=7cm,height=5cm]{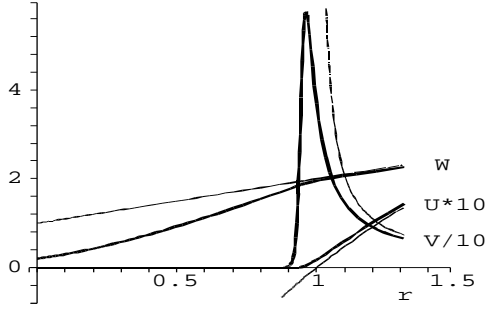}
\caption{Black hole functions $U(r)$, $V(r)$ and $W(r)$, scaled by factors $10$, (bold lines) 
for $\mub=0.1$, compared to the Schwarzschild solution (thin lines; the part $V<0$ for $r<1$ is not shown). 
Inside the horizon, $U$ and $V$ decay very rapidly. Since they remain positive,
time keeps its role in the interior.}
\end{figure}

{\it Width of the brick wall}.
The transition layer $\xi={\cal O}(1)$ acts like 't Hooft's brick wall,~\cite{tHooft}
of characteristic width $\ell_\star=\eta r_h$.
Comparing to the Planck length $\ell_P=\sqrt{\hbar G/c^3}$,  we get
\BEQ
\frac{\ell_\ast}{\ell_{\rm P}}=\frac{0.977\,\,10^{-9}}{1+0.019\log(M/M_\odot)}
\,\frac{M^3}{M_\odot^3}.
\EEQ
If quantum physics sets in at the Planck scale, our approach
makes sense only for $ M>10^3\,\,M_\odot$.

{\it Motion of test particles}.
For RTG with a negative cosmological constant, ~\cite{LogunovBook} it was claimed that an incoming spherical 
shell of matter is scattered off from a black hole, a counter-intuitive finding.
Let us reconsider this issue. The motion of a test body occurs along a geodesic
\BEQ \frac{\d v^\mu}{\d s}+\Gamma^\mu_{\nu\rho}v^\nu v^\rho=0,
\qquad v^\mu=\frac{\d x^\mu}{\d s}. 
\EEQ
For spherical shells of in-falling matter one needs $\Gamma^{0}_{01}=U'/(2U)$.
This brings $\d t/\d s=v^0=1/({C_i}{U})$,  
for some $C_i$. Solving $v^1=\d r/\d s$ from $g_\mn v^\mu v^\nu=1$, we then get
$\label{rdot=} {\d r}/{\d t}=({\d s}/{\d t})({\d r}/{\d s})=-c\sqrt{U(1-C_i^2U)/V}.$
We can now fix $C_i$ at the initial position $r=r_i$, where the spherical shell is assumed
to have a speed $\d r_i/\d t=v_i=\beta_ic\sqrt{U_i/V_i}$, viz.
 $C_i=\sqrt{(1-\beta_i^2)/U_i}$, with $|\beta_i|\le1$.
The differential proper time $\d\tau=\sqrt{U}\d t$ and length $\d\ell=\sqrt{V}\d r$ 
bring in the particle's rest frame $\d\ell/\d\tau=\sqrt{V/U}\d r/\d t$, yielding
\BEQ \frac{\d \ell}{\d \tau}=-c\sqrt{1-\frac{U(r(\tau))}{U(r_i)}(1-\beta_i^2)}.
\EEQ
The extreme case is when $\beta_i=0$ at $r_i=\infty$, ${\d \ell}/{\d \tau}
=-c\sqrt{1-U}.$
To have $|{\d \ell}/{\d \tau}|< c$, it thus suffices that $0<U\le1$, which is the case.
Near the horizon, $|\d\ell/\d\tau|$ is almost equal to $c$ and 
the more the shell penetrates the interior, the closer its speed gets to $c$.
For an outside observer, the time to see it hit the center of the hole,
$T=\int\d r/|\dot r|$ equals $(r_h/c)\int_0^1\d r\exp[\half(v-u)]$.
It is finite and predominantly comes from the horizon, 
$T=r_h/c\mub^2= 2.74\times10^{32}M/M_\odot$ yr.

The approaches ~\cite{LogunovBook,LogInflaton,BabakGrishchuk2} have a similar
a black hole.
While \cite{BabakGrishchuk2} properly has $U'(0)=0$,
in Logunov's case one has $\bar\mu^2<0$,
so $w'=\half\alpha <0$ in the interior. This seems to solve the paradox of 
``matter reflected by the black hole'':
In-falling matter just enters, but the Logunov coordinate $x=\exp (w)-1$ is non-monotonic 
($x'<0$ in the interior). However, the situation is more severe: 
For $\alpha<0$, the theory does not allow a solution with $u'(0)=0$, depriving that 
theory of a proper black hole. This condition can neither be obeyed in GTR: 
{\it If the central mass is slightly smeared, 
the Schwarzschild black hole cannot obey energy-momentum conservation in GTR.}

{\it Cosmology}. 
Starting from the FLRW metric,
the harmonic condition brings two relations: $U\sim V^3$ and
$k=0$: Minkowski space filled homogeneously with matter remains flat
~\cite{LogunovBook}. We may thus put $U=a^6(t)/a_\ast ^4$, $V=a^2(t)$. 
Going from cosmic time $t$ to conformal time $\tau=\int a^3a_\ast ^{-2}\d t$
yields the familiar Einstein equations, extended by $\Lambda_\bi$ terms, 
\BEQ \label{Friedman}\frac{\dot a^2}{a^2c^2}&=&\frac{8\pi G}{3c^4}\rho+
\frac{\Lambda}{3}-\frac{\Lambda_\bi}{2a^2}
+\frac{\Lambda_\bi a_\ast ^4}{6a^6},\quad \\ 
\frac{\ddot a}{a\,c^2}&=&-\frac{4\pi G}{3c^4}\,(\rho+3p)
+\frac{\Lambda }{3}-\frac{\Lambda_\bi }{3}\frac{a_\ast ^4}{a^6}.\nn
\EEQ 
The first is the modified Friedman equation, 
the second corresponds to the first law $\d(\rho_{\rm tot}a^3)=-p_{\rm tot}\d a^3$ 
provided we define $\rho_{\rm tot}=\rho+\rho_\Lambda+\rho_2+\rho_6$ 
and $p_{\rm tot}=p-\rho_\Lambda-\frac{1}{3}\rho_2+\rho_6$, with
$\rho_2=-3\rho_\bi/2a^2$ and $\rho_6= \rho_\bi a_\ast^4/2a^6$. 
Note that $\rho_2$ acts as a positive curvature term.

The scale factor has an absolute meaning. 
If we assume that $a\gg1$ and $a\gg a_\ast^{2/3}$, Eq. (\ref{Friedman}) 
just coincides with the $\Lambda$CDM model (cosmological constant plus cold dark matter), 
that gives the best fit of the observations~\cite{ThemAgain}\cite{WMAP}.
The $\rho_2$ term allows a positive curvature-type contribution.
At large times, there is the exponential growth
$a(\tau)=C \exp(H_\infty\tau)$ with $H_\infty=c\sqrt{\Lambda/3}$.
In cosmic time this reads $a(t)=a_\ast^{2/3}[3H_\infty(t_0-t)]^{-1/3}$, where $t_0$
is ``the end of time'', the moment where the scale factor has become infinite.
The minimal scale factor is zero: in this theory a big bang can occur since $\rho_\bi>0$.
Without including an inflaton field, Eq. (\ref{Friedman}) yields an initial growth
of the expansion  $a=(a_\ast^2c\,\tau\sqrt{3\Lambda_\bi/2} )^{1/3}$. 
In cosmic time this reads $a=a_1\,\exp(ct\sqrt{\Lambda_\bi/6})$, 
i. e., a certain inflation scenario starting at $t=-\infty$.

Also in RTG the gravitational energy precisely compensates 
the other energy contributions at all times.
The vacuum energy also vanishes: 
In empty space, the cosmological constant energy $\rho_\Lambda$ cancels
 the $\rho_\bi$ terms, due to Eq. (\ref{omomom}). See Eq. (\ref{Einst=}) 
for $g_\mn=\gamma_\mn$.

{\it In conclusion},  we have first written the Einstein equation in a form
that involves the gravitational energy momentum tensor.  
An underlying Minkowski space is needed, in which gravitation is a field. 
The metric tensor is a way to deal with it,
but the equations for the field itself exist too, see Eq. (\ref{MaxEinEq=}).
For flat cosmology it follows that the total energy vanishes.

Next we have broken general coordinate invariance by going to the bimetric 
theory of Logunov, called Relativistic Theory of Gravitation.
We have shown that the choice of a positive bimetric constant
allows to regularize the interior of the Schwarzschild black hole:
time keeps its standard role and escape is, in principle, possible.
While neither the Schwarzschild nor the Logunov black hole survives smearing
of the central mass by a tiny pressure in the equation of state, ours does.
Our modification of the Einstein equations involves the cosmological constant,
so it is of Hubble size, immaterial for solar problems.
In cosmology, the theory directly leads to the $\Lambda$CDM model, while it could 
accommodate a positive curvature-like term. At short times, there is a form of inflation. 
The gravitational energy exactly compensates the material energy.
The zero point energy vanishes (``again''), though the cosmological constant 
is finite and positive: It is canceled by the bimetric terms.

Euclidean space, a special case of Riemann geometry, 
seems to be invoked by Nature, at least far away from bodies and in cosmology.
Our approach supports the following space-time interpretation: 
curvature is a geometric description of the gravitational field in flat space.
Clusters of galaxies do not move away from each other, 
but the speed of light changes with cosmic time, $\d r/\d t=[a(t)/a_\ast]^2c$,
while the conformal speed is $\d r/\d\tau=c/a(\tau)$ as usual.

An empirical way to establish the Minkowski metric is to
present the Einstein equations as
$(c^4/8\pi G)R_\mn-T_\mn+\half g_\mn T+\rho_\Lambda g_\mn=\rho_\bi\gamma_\mn$,
and to measure the left hand side,
which in the geometric view is considered to consist of
curved space properties alone.~\cite{LogunovBook}

As in the standard model of elementary particles, 
the separation of curved space into flat space 
and the gravitational field has the following implication: 
the quantum version of RTG -- if it exists -- will involve 
quantization of fields, but not of space.

\vspace{0.5cm}

Finally we answer the question posed in the title.
The field theoretic approach to gravitation is by itself 
equivalent to a curved space description, so both 
views apply, describing the same physics from a different angle.
 But when the theory is extended to the 
relativistic theory of gravitation, the bimetrism forces to describe 
the Minkowski metric separately, and then we see 
it as most natural to view gravitation as a field in flat space,
which is Maxwell's view.

Topics such as a realistic equation of state for black holes 
and classical tunneling of its radiation, regularization of other singularities,
as well as aspects of the inflation and of inhomogeneous cosmology are under study.

\acknowledgments
Discussion with Martin Nieuwenhuizen and Armen Allahverdyan
 is gratefully remembered.


\end{document}